\documentclass[twocolumn,showpacs,preprintnumbers,amsmath,amssymb]{revtex4}

\usepackage{graphicx}
\usepackage{amssymb}
\usepackage{dcolumn}
\usepackage{color}

\newcommand {\be}{\begin{equation}}
\newcommand {\ee}{\end{equation}}
\newcommand {\bea}{\begin{eqnarray}}
\newcommand {\eea}{\end{eqnarray}}
\newcommand {\FIG}[1]{Fig. \ref{#1}}
\newcommand {\EQ}[1]{Eq. (\ref{#1})}

\newcommand {\TAB}[1]{Table \ref{#1}}

\newcommand {\PRE}[1]{{Phys. Rev. E} {\bf {#1}}}

\newcommand {\NAT}[1]{{Nature} {\bf {#1}}}

\begin{document}

\draft
\title{Statistical properties of sampled networks by random walks}
\author{Sooyeon Yoon, Sungmin Lee}
\author{ Soon-Hyung Yook}\email{syook@khu.ac.kr}
\author{Yup Kim} \email{ykim@khu.ac.kr}
\affiliation{Department of Physics and Research Institute for Basic
Sciences, Kyung Hee University, Seoul 130-701, Korea}
\date{\today}
\begin{abstract}
We study the statistical properties of the sampled networks by a
random walker. We compare topological properties of the sampled
networks such as degree distribution, degree-degree correlation, and
clustering coefficient with those of the original networks. From the
numerical results, we find that most of topological properties of
the sampled networks are almost the same as those of the original
networks for $\gamma \lesssim 3$. In contrast, we find that the
degree distribution exponent of the sampled networks for $\gamma>3$
somewhat deviates from that of the original networks when the ratio
of the sampled network size to the original network size becomes
smaller. We also apply the sampling method to various real networks
such as collaboration of movie actor, world wide web, and
peer-to-peer networks. All topological properties of the sampled
networks show the essentially same as the original real networks.

\end{abstract}

 \maketitle

Since the concept of complex network \cite{cn} came into the
limelight, many physically meaningful analyses for the complex
networks in real world have emerged. The examples of such studies
include the protein-protein interaction networks (PIN)
\cite{syook03}, world wide web (WWW) \cite{albert99}, email network
\cite{ebel02}, etc. The empirical data or the information of the
real networks can be collected in various ways, for example, the
traceroutes for the Internet \cite{internet} and high throughput
experiments for protein interaction map \cite{uetz00}. Thus, it is
natural assumption that the empirical data can be incomplete due to
various reasons which include some limitations of the experiments
and experimental errors or biases. As a result, many real networks
which have been intensively studied so far can be regarded as
sampled networks. Moreover, several studies have shown that the
dynamical properties on the networks can be significantly affected
by the underlying topology \cite{rw,dcpLee}. Therefore, it is very
important and interesting to study the topological differences
between the sampled networks and the whole networks.

Recently, several sampling methods such as random node sampling
\cite{Stumpt,LeeS}, random link sampling, and snowball sampling were
studied \cite{LeeS}. The random node sampling is the simplest method
in which the sampled network consists of randomly selected nodes
with a given probability $p$ and the original links between the
selected nodes. On the other hands, in the random link sampling, the
links are randomly selected and the nodes connected to the selected
links are kept. These two random sampling methods have been used to
study the statistical survey in some social networks. In the random
sampling method, however, many important nodes such as hubs cannot
be sampled due to the even selection probability. Some recent
studies show that some networks such as PIN, the topological
properties of randomly sampled networks significantly deviate from
those of the original networks \cite{Stumpt,LeeS}. The idea of the
snowball sampling method \cite{LeeS,egoNewman} is similar to the
breath-first search algorithm \cite{bNewman,breath}. In the snowball
sampling method all the nodes directly connected to the randomly
chosen starting node are selected. Then all the nodes linked to
those selected nodes in the last step are selected hierarchically.
This process continues until the sampled network has the desired
number of nodes \cite{LeeS}. Previous studies showed that the
topological properties of the sampled networks closely depend on the
sampling methods \cite{LeeS}.

In this paper, we focus on the effect of the weighted sampling on
the topological properties of sampled networks. In order to assign
nontrivial weight to each node, we first note the structure of the
real networks. Many real networks are known to be the scale-free
networks in which the degree distribution follows the power-law
\cite{cn},
\begin{equation}
P(k)\sim k^{-\gamma}~. \label{pkd}
\end{equation}
Moreover, the probability $p_v(k)$ that a random walker (RW) visits
a node of degree $k$ \cite{rw} is given by
\begin{equation}
p_v(k)\sim k ~.\label{pbrw}
\end{equation}
The degree $k$ causes uneven probability of finding a node by a RW
on the heterogeneous networks. Thus, by using the RW for sampling we
can assign automatically the nontrivial weight to each node which is
proportional to the degree of the node. Due to the uneven visiting
probability, the nodes having the large degree, i.e., topologically
important parts, can be easily found regardless of the starting
position of the RW. Therefore, we expect that the sampling by the RW
can provide more effective way to obtain the sub-networks which have
almost the same statistical properties as the original one.
Furthermore, we also study the effects of the heterogeneity of the
original networks on the RW sampling method (RWSM) by changing
$\gamma$. This weighted sampling method is also shown to be
successfully applied to obtain the important informations of many
real networks such as actor networks, WWW, and peer-to-peer (P2P)
networks. Therefore, we expect that this study can provide a better
insight to understand important properties of the real networks and
offer a systematic approach to the sampling of networks with various
$\gamma$.


We now introduce the RWSM. First, we generate orignal scale-free
networks (SFN) by use of the static model in Ref. \cite{staticGoh}
from which various sizes of sub-networks are sampled. The size or
number $N_o$ of nodes of the original network with each $\gamma$
is set to be $N_o=10^6$. The typical values of $\gamma$ used in
the simulations are $\gamma=2.23, 2.51, 3.05, 3.40$, and $4.2$. We
set the average degree $\langle k \rangle=4$ for each network.
After the preparation of original networks, a RW is placed at a
randomly chosen node and moves until it visits $N_s$ distinct nodes.
Then we construct sub-networks with these $N_s$ visited nodes and
the links which connect any pair of nodes among the $N_s$ visited
nodes in the original network. We use $N_s=10^3, 10^4, 2\times 10^4,
4\times 10^4, 6\times 10^4, 8\times 10^4, 10^5$, and $1,2,3, \cdots,
9 \times 10^5$.

\begin{figure}[h]
\includegraphics[width=9cm]{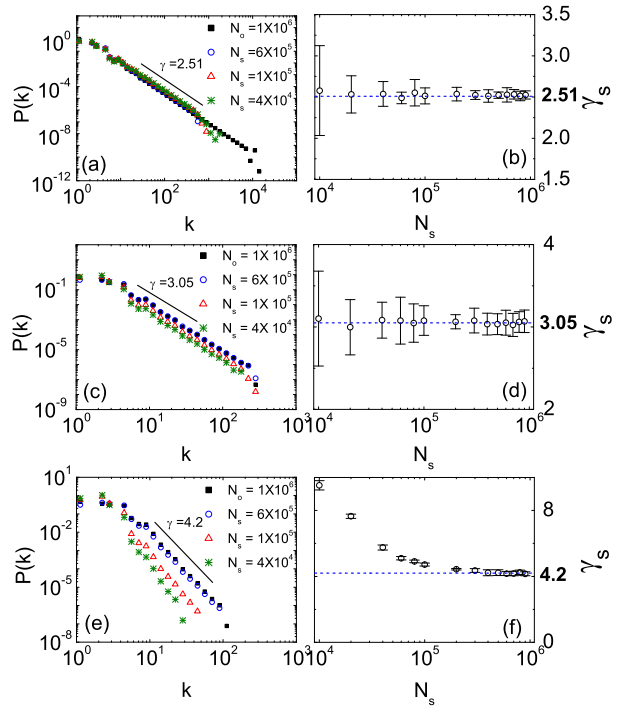}
\vspace{5mm} \caption{(Color online) Degree distributions for
sampled networks of static scale-free networks with (a)
$\gamma=2.51$, (c) 3.05, and (e) 4.2. Degree exponents $\gamma_s$
for the sampled networks extracted from the original network for the
network size $N_o=10^6$ with (b) $\gamma=2.51$, (d) $\gamma=3.05$,
and (f) $\gamma=4.2$. The slopes of solid lines in (a,c,e) and the
values of the dashed lines in (b,d,f) are the degree exponents of
the original networks.}\label{pksf}
\end{figure}

The degree distribution is one of the most important measure for
the heterogeneity of networks \cite{cn}. In \FIG{pksf}, we compare
the degree distributions of the sampled networks  to those of the
original networks for various $\gamma$. We find that the degree
distribution of the sampled network also satisfies the power-law,
$P(k)\sim k^{-\gamma_s}$.

Especially, from the data in Figs. \ref{pksf} (a)-(d) we find that
the $\gamma_s$ of the sampled networks with $N_s/N_o \ge 0.01$ is
nearly equal to $\gamma$ of the original network, even though the
$\gamma_s$ for the small $N_s$ has relatively larger error bar. It
shows that the sampling method by RW does not change the
heterogeneity in degree for networks with $2<\gamma\lesssim 3$.
Since most of the real networks have $2<\gamma<3$ \cite{cn}, this
result is practically important.
We summarize the obtained $\gamma_s$'s for various $N_s$'s and
$\gamma$'s in \TAB{t1}.

In contrast to the case $\gamma \lesssim 3$, $\gamma_s$ for $\gamma
> 3$ slightly deviates from the $\gamma$ of the original networks if
$N_s /N_o \lesssim 0.1$. (See the data for $\gamma=4.2$ in Figs.
\ref{pksf} (e) and (f) or in \TAB{t1}.) Numerically we find that
$\gamma_s$ is nearly equal to the original $\gamma$ for $N_s/N_o>
0.1$ when $\gamma \lesssim 4.2$. Of course one can expect the
substantial deviation  of $\gamma_s$ from $\gamma$ as $\gamma$
increases further from $\gamma=4.2$.

This $\gamma$ dependent behavior of $P(k)$ can be understood from
Eqs. (\ref{pkd}) and (\ref{pbrw}). Equation (\ref{pkd}) indicates
that $\langle k^2 \rangle$ diverges with finite $\left<k\right>$ for
$\gamma \le 3$. This implies that the topology of a network has
several dominant hubs which have extraordinary large number of
degrees when $\gamma<3$. Since the probability of visitation of the
RW follows \EQ{pbrw}, the RW can more effectively find the central
part of the network around the hubs when $\gamma<3$. Thus the
sampled networks can inherit easily the topological properties of
the original networks.

\begin{figure}[h]
\includegraphics[width=8cm]{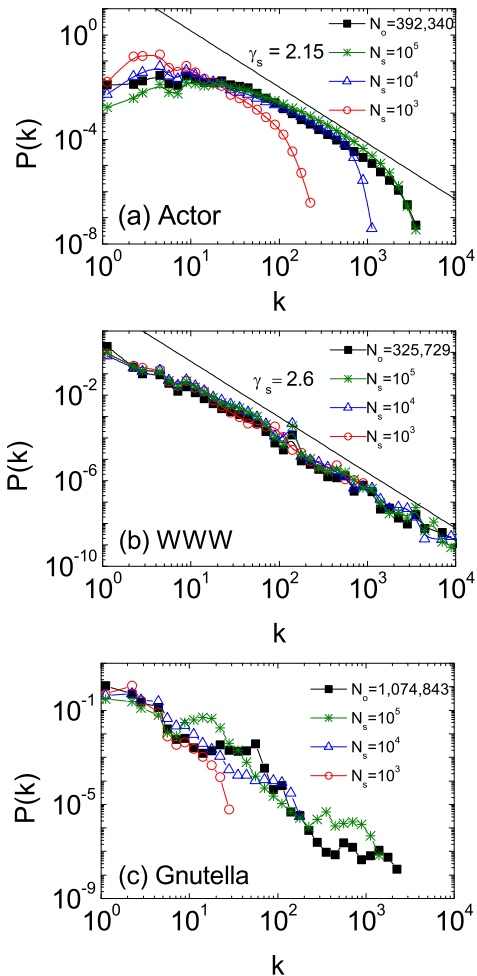}
\vspace{5mm} \caption{(Color online) Degree distributions for
sampled networks of three real networks. (a) Collaboration network
of movie actors ($N_o=392,340, \gamma=2.2$) \cite{actor}, (b) WWW
($N_o=325,729, \gamma=2.6$) \cite{albert99}, and (c) Gnutella
($N_o=1,074,843$) \cite{gnutella}. The slopes of the solid lines in
(a) and (b) are the values of degree exponents obtained from the
simple linear fitting for degree distributions of the sampled
networks.}\label{pkr}
\end{figure}

RWSM is also applied to real networks. In \FIG{pkr}, we show the
$P(k)$ of the actor network \cite{actor}, the WWW \cite{albert99},
and the P2P networks (Gnutella) \cite{gnutella}. The number of nodes
in the original real networks are $N_o=392340,~ 325729$, and
$1074843$ for the actor network, the WWW, and the Gnutella,
respectively. The degree distributions for the actor network and the
WWW follow the power-law with $\gamma=2.2$ (actor) \cite{actor} and
$\gamma=2.6$ (WWW) \cite{albert99}. The data in \FIG{pkr} (a) shows
that $P(k)$ of the sampled actor network follows the power-law with
$\gamma_s=2.15$ for $N_s>10^3$. This value of $\gamma_s$ is quite
close to $\gamma=2.2$. In contrast $\gamma_s$ seems to deviate from
$\gamma$ of the original network for small $N_s (= 10^3$). However,
the degree exponent $\gamma_s$ for $N_s=10^3$ still has almost the
same value with that of the original network over one decade
($k=10\sim 100$). In the case of the WWW, the $\gamma_s$ of the
sampled networks well agrees with $\gamma\simeq 2.6$ even for small
$N_s(=10^3)$ (see \FIG{pkr}(b)). For the Gnutella, $P(k)$ of the
original network does not follow the simple power-law (\ref{pkd}).
However, as one can see in Fig. \ref{pkr} (c), the Gnutella network
also has big hubs which cause high heterogeneity in degree, and the
sampled networks nearly show the same degree distribution as the
original one. These results also provide the evidence that the nodes
with large degrees play an important role in the RWSM.

\begin{figure}[h]
\vspace{5mm}
\includegraphics[width=9cm]{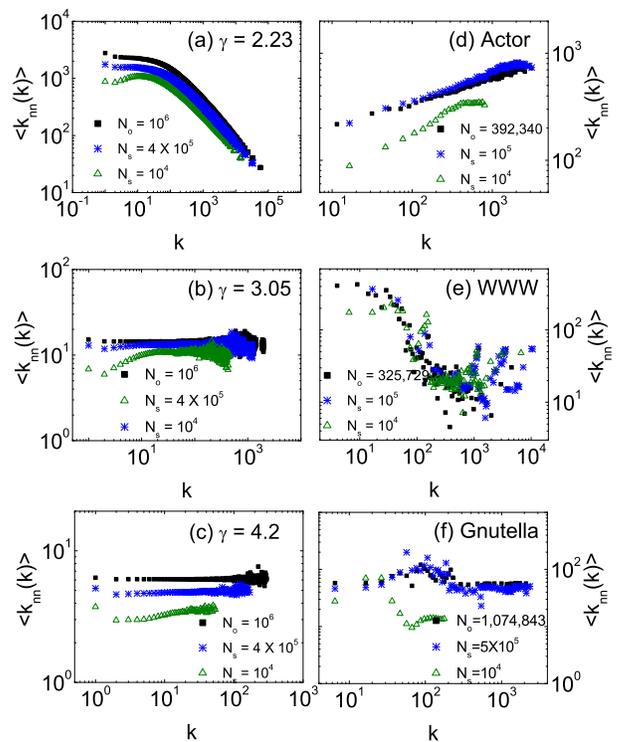}
\caption{(Color online) Distributions of $\langle k_{nn}\rangle$ for
sub-networks extracted from the original networks with (a)
$\gamma=2.23$, (b) $\gamma=3.05$, and (c) $\gamma=4.2$. (d)
Collaboration network of movie actors. (e) WWW. (f)
Gnutella.}\label{corr}
\end{figure}

Another important measure to characterize the topological properties
of the complex network is the degree-degree correlation. Many
interesting topological properties such as the self-similarity
\cite{ss} can be affected by the degree-degree correlation. The
degree-degree correlation can be characterized by $\langle k_{nn}
(k)\rangle$, the average degree of the nearest-neighbors of nodes
with degree $k$ \cite{corrPastor, corrNewman}. If the $\langle
k_{nn}(k)\rangle$ increases (decreases), the network is
characterized as assortative (disassortative) mixing. As shown in
\FIG{corr} (a), for the static SFN with $2<\gamma<3$ the original
network and the sampled networks all show the disassortative mixing.
This can be explained by the dynamical properties of RWs on complex
networks. In the networks showing disassortative mixing, the RW on a
hub should go through a node of small $k$ to move to another hub.
Thus, many nodes having small $k$ can be connected to the hubs in
the sampled networks and the sampled networks remain disassortative.
If the networks have neutral degree correlation, then the networks
sampled by the RW also show neutral degree correlation. (See Figs.
\ref{corr} (b) and (c).) In Figs. \ref{corr} (d)-(f), we plot
$\langle k_{nn} (k)\rangle$ of real networks. $\langle k_{nn}
(k)\rangle$'s of the sampled networks show the same degree
correlations as those of the original networks. As shown in Figs.
\ref{corr} (d)-(f), the degree correlations are assortative,
disassortative, and neutral for the actor, WWW, and Gnutella
networks, respectively.

We also measure a clustering coefficient of the sampled networks.
The clustering coefficient $C_i$ of a node $i$ is defined by
\begin{equation}
C_i = \frac {2y_i}{k_i (k_i -1)}~,
\end{equation}
where $k_i$ is the degree of node $i$ and $y_i$ is the number of
connections between its nearest neighbors \cite{cn}. $C_i$
physically means the fraction of connected pairs among pairs of
node $i$'s neighbors. $C_i$ is one if all neighbors are completely
connected, whereas $C_i$ becomes zero on a infinite-sized random
network \cite{cn}.

\begin{figure}[h]
\vspace{5mm}
\includegraphics[width=9cm]{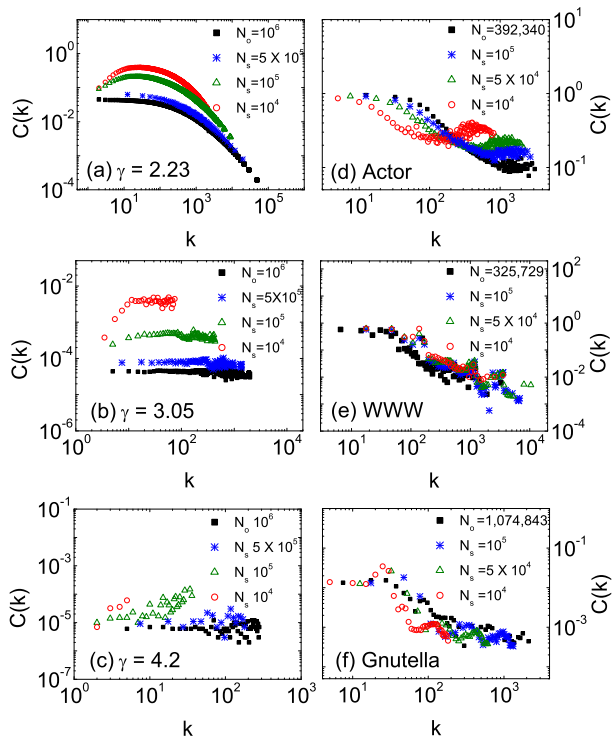}
\caption{(Color online) $C(k)$ for sub-networks from the original
networks with (a) $\gamma=2.23$, (b) $\gamma=3.05$, and (c)
$\gamma=4.2$. (d) Collaboration network of movie actors. (e) WWW.
(f) Gnutella.}\label{clus}
\end{figure}

In Fig. \ref{clus}, we plot the clustering coefficient $C(k)$
against degree $k$. $C(k)$ is known to reflect the modular structure
of networks \cite{corrPastor,corre}. $C(k)$ does not depend on $k$
if the network does not have any well defined hierarchical modules
\cite{corrPastor,corre}. As shown in Fig. \ref{clus}, $C(k)$ of both
the original networks and the sampled networks shows a tendency to
decrease with increasing $k$ for SFN with $\gamma=2.23$ and real
networks. (See Figs. \ref{clus} (a) and (d)-(f)). This implies that
the sampled networks have the same modular structure with original
networks. On the other hand, the topology of networks with $\gamma
\gg 3$ resembles closely the random graph, thus $C(k)$ does not
depend on the degree $k$ \cite{corre}. The dependence of $C(k)$ on
$k$ for the sampled SFNs with $\gamma\ge3$ is also nearly the same
as the original SFNs. (See Figs. \ref{clus} (b) and (c).)

We study the topological properties of sampled networks by RWSM with
SFN and several real networks. From the numerical simulations, we
find that the $P(k)$ of the sampled network follows the power-law,
$P(k)\sim k^{-\gamma_s}$. We also find that the $\gamma_s\simeq
\gamma$ for all $N_s$ when $2< \gamma\lesssim 3$. Even though
$\gamma_s$ somewhat increases as decreasing $N_s$ for $\gamma>3$,
the $\gamma_s$'s with $N_s/N_o\gtrsim0.1$ still follow the original
one. We also study the degree-degree correlation and clustering
coefficient by measuring $\langle k_{nn}(k)\rangle$ and $C(k)$. The
sampled networks have the same degree correlation and modular
structure with the original networks for all values of $\gamma$. The
RWSM is also applied to the actor, WWW, and Gnutella networks. By
measuring $P(k)$, $\langle k_{nn}(k)\rangle$, and $C(k)$, we confirm
that the topological properties of the sampled networks are well
maintained after sampling and the RWSM is efficient sampling method
for the real networks.

The $\gamma$ dependent behavior of the sampled networks can be
understood from the dynamical property of a RW. Since most of the
networks in the real world have $2<\gamma<3$, the results imply very
important meaning in practice. Based on our results, we expect that
if we obtain the empirical data by weighted sampling in which the
weight is proportional to the degree, then the sampled networks can
share the same topological properties with the whole network.
Especially, the weighted sampling method becomes very efficient as
the heterogeneity of networks increases. At the same time, we also
expect that our study can provide a systematic way to extract
sub-networks from the empirical data and to study various dynamical
properties of the real networks \cite{lee}.

\begin{acknowledgments}
This work is supported by grant No. R01-2006-000-10470-0 from the
Basic Research Program of the Korea Science \& Engineering
Foundation.
\end{acknowledgments}

\begin{widetext}
\begin{center}
\begin{table}
\caption{The changes of the degree distribution exponents
$\gamma_s$ depending on sampled network size $N_s$. $\gamma$'s are
the degree exponents of the original network with $N_o=10^6$.}
\vspace{5mm}
\begin{tabular} { c | c  c c c c c c c c c c }
\hline\hline  &&&&&$N_s$ ($\times 10^6$) \\ $\gamma$ & $0.8$ &
$0.6$ & $0.4$ & $0.2$ & $0.1$
& $0.08$ & $0.06$ & $0.04$ & $0.02$ & $0.01$ \\
\hline $2.23(5)$ & 2.23(3) & 2.24(3) & 2.24(2)& 2.24(3) & 2.3(1) &
2.2(2) & 2.3(2) &
2.3(3) & 2.3(5) & 2.3(5) \\
$2.51(7)$ & 2.51(6) & 2.53(8) & 2.51(8)& 2.54(8) & 2.5(1) & 2.6(2)
& 2.49(7) &
2.5(1) & 2.5(1) & 2.6(5) \\
$3.05(9)$ & 3.1(1) & 3.1(2) & 3.0(1)& 3.06(9) & 3.1(2) & 3.1(2) &
3.1(3) &
3.1(2) & 3.0(3) & 3.1(6) \\
$3.40(8)$ & 3.37(7) & 3.40(9) & 3.4(1) & 3.4(1) & 3.4(2) & 3.5(3)
& 3.4(2) &
3.7(4) & 3.8(4) & 4.4(3) \\
$4.2(1)$ & 4.2(1) & 4.2(1) & 4.2(2)& 4.44(5) & 4.71(9) &
4.91(8) & 5.1(1) &5.8(2) & 7.7(1) & 9.5(3) \\
\hline\hline
\end{tabular}
\label{t1}
\end{table}
\end{center}
\end{widetext}

\end{document}